\documentclass[12pt]{article}
\usepackage{amsfonts}
\begin{document}
\footskip 1cm
\parskip=1ex

% the following command must be inserted (uncommented) when there are
% sections.
%\renewcommand\theequation{\thesection.\arabic{equation}}

% almost universal commands for equations and references
\newcommand{\eqn}[1]{(\ref{#1})}
\newcommand{\be}{\begin{equation}}
\newcommand{\ee}{\end{equation}}
\newcommand{\bea}{\begin{eqnarray}}
\newcommand{\eea}{\end{eqnarray}}
\newcommand{\bean}{\begin{eqnarray*}}
\newcommand{\eean}{\end{eqnarray*}}
\newcommand{\nn}{\nonumber}

% footnotes with symbols instead of numbers
\renewcommand{\thefootnote}{\fnsymbol{footnote}}

% Miscellaneous mathematical symbols

%%% Operator names:
\newcommand{\opname}[1]{\mathop{\rm #1}\nolimits} %% like \det, \ker,...
\newcommand{\Tr}{\opname{Tr}} %% trace of operator
\newcommand{\tr}{\opname{tr}} %% trace of matrix

\def\slash#1{\rlap/#1}
\newcommand{\alg}{{\cal A}} %% an algebra
\newcommand{\Aslash}{A\mkern-11.5mu/\,} %% gauge potential
\newcommand{\barox}{\mathrel{\overline\otimes}} %% graded tensor product
\newcommand{\bra}[1]{\langle{#1}\vert} %% bra vector
\newcommand{\stroke}{\mathbin{\vert}} %% (for `\braket' and such)
\newcommand{\braket}[2]{\langle#1\stroke#2\rangle} %% Dirac bracket
\newcommand{\del}{\partial} %% abbreviation for  \partial
\newcommand{\delslash}{{\partial\mkern-9mu/}} %% usual Dirac operator
\newcommand{\Dslash}{D\mkern-11.5mu/\,} %% generalized Dirac operator
\newcommand{\eps}{\varepsilon} %% abbreviation for  \epsilon
\newcommand{\Hilb}{{\cal H}} %% Hilbert space
\newcommand{\ket}[1]{\vert{#1}\rangle} %% ket vector
\newcommand{\row}[3]{{#1}_{#2},\dots,{#1}_{#3}} %% list:  a_1,...,a_n
\newcommand{\thalf}{{\textstyle\frac{1}{2}}} %% small fraction  1/2
\newcommand{\tihalf}{{\textstyle\frac{i}{2}}} %% small fraction  i/2
\newcommand{\tquarter}{{\textstyle\frac{1}{4}}} %% small fraction  1/4
\newcommand{\vev}[1]{\langle#1\rangle} %% vacuum expectation value
\def\<#1,#2>{\langle#1\stroke#2\rangle} %% inner product (Dirac not'n)

\newcommand{\complex}{{\mathbb C}} %% complex numbers
\newcommand{\quater}{{\mathbb H}} %% quaternions
\newcommand{\unit}{{\mathbb I}} %% identity matrix
\newcommand{\natu}{{\mathbb N}} %% nonnegative integers
\newcommand{\real}{{\mathbb R}} %% real numbers
\newcommand{\inte}{{\mathbb Z}} %% all integers

% A macro to raise things. Used in math and journal macros.
\def\up#1{\leavevmode \raise.16ex\hbox{#1}}
%journal references
\newcommand{\npb}[3]{{\sl Nucl. Phys. }{\bf B#1} \up(#2\up) #3}
\newcommand{\plb}[3]{{\sl Phys. Lett. }{\bf #1B} \up(#2\up) #3}
\newcommand{\revmp}[3]{{\sl Rev. Mod. Phys. }{\bf #1} \up(#2\up) #3}
\newcommand{\sovj}[3]{{\sl Sov. J. Nucl. Phys. }{\bf #1} \up(#2\up) #3}
\newcommand{\jetp}[3]{{\sl Sov. Phys. JETP }{\bf #1} \up(#2\up) #3}
\newcommand{\rmp}[3]{{\sl Rev. Mod. Phys. }{\bf #1} \up(#2\up) #3}
\newcommand{\prd}[3]{{\sl Phys. Rev. }{\bf D#1} \up(#2\up) #3}
\newcommand{\ijmpa}[3]{{\sl Int. J. Mod. Phys. }{\bf A#1} \up(#2\up) #3}
\newcommand{\mpla}[3]{{\sl Mod. Phys. Lett. }{\bf A#1} \up(#2\up) #3}
\newcommand{\prl}[3]{{\sl Phys. Rev. Lett. }{\bf #1} \up(#2\up) #3}
\newcommand{\physrep}[3]{{\sl Phys. Rep. }{\bf #1} \up(#2\up) #3}
\newcommand{\jgp}[3]{{\sl J. Geom. Phys.}{\bf #1} \up(#2\up) #3}
\newcommand{\journal}[4]{{\sl #1 }{\bf #2} \up(#3\up) #4}

% some hyphenations it is better to have
\hyphenation{geo-me-try}
\hyphenation{Liz-zi}

% some frequently used abbreviations

\def\cstars{$C^*$-algebras }
\def\cstar{$C^*$-algebra }
\def\iff{\Leftrightarrow}

\thispagestyle{empty}
\setcounter{page}{0}
\begin{flushright}
DSF/19-99\\
\hfill hep-th/9906122\\
May 1999
\end{flushright}
\vspace{.5cm}
\begin{center}{\Large \bf Noncommutative Geometry, Strings and Duality }
\end{center}
\vspace{.5cm}
\centerline{\large Fedele Lizzi\footnote{fedele.lizzi@na.infn.it}
}
\vspace{.25cm}
\begin{center}

~\\ {\it Dipartimento di Scienze Fisiche, Universit\`a di Napoli Federico
II, \\ and INFN, Sezione di Napoli\\ Mostra d' Oltremare pad. 20, I-80125,
Napoli, Italy.}
\end{center}
\vspace{1.5cm}
\begin{center}
Talk delivered at the Arbeitstagung:\\ {\sl The standard Model of
Elementary particle Physics,\\
 Mathematical and Geometrical Aspects},\\
Hesselberg, March 14-19 1999.
\end{center}
\begin{abstract}
In this talk, based on work done in collaboration with G.~Landi and
R.J~Szabo, I will review how string theory can be considered as a
noncommutative geometry based on an algebra of vertex operators. The
spectral triple of strings is introduced, and some of the string
symmetries, notably target space duality, are discussed in this framework.
\end{abstract}
\newpage
%%% --------------------------------------------------------------
It is a common belief that at distances of the order of Planck's length,
where neither quantum mechanics nor general relativity can be considered
perturbations of classical physics, a change in the very structure of
spacetime will be necessary, and {\em classical geometry} will no longer be
the appropriate tool. A suitable candidate for the mathematics:  is {\it
Noncommutative Geometry}, \cite{book}, a theory which substitutes the study
of {\em classical} notions such as point, line etc. with the study of the
algebras defined on Hausdorff topological spaces, with the obvious
generalization given by noncommutative algebras. I will not introduce
Noncommutative Geometry here, there are excellent reviews, and it is
covered in several of the contributions to this Arbeitstagung.

On the physics side, a candidate to describe the theory of physical
interactions at the Planck length is {\em String Theory} \cite{GSW}. An
introduction of strings will also be a task too formidable for this short
talk, but the basic motivation behind string theory as a fundamental theory
of spacetime, historically has been the presence in the spectrum of the
theory of a spin two massless particle, identified with the graviton.
String theory is a two-dimensional theory in which spacetime coordinates
are fields on a 2-d surface, the world-sheet of the string. Interactions,
that is joining and splitting of strings, is described by higher genus
surfaces. At very high energies (of the order of Planck's energy),
``strange things" begin to happen {\em $\infty$-genus surfaces, duality,
branes whose coordinates are (noncommuting) matrices...}

In this talk I will try to construct the {\em Noncommutative Geometry of
String Theory}. The key idea is due to Fr\"ohlich and Gaw\c edzki
\cite{FG}, and in this talk I will mainly sketch the developments of
\cite{lizziszaboprl}-\cite{Corfu}, to which I refer for further details
and references.

String theory is described by a conformally invariant field theory on a
2-dimensional surface, the world--sheet, which is to be interpreted as the
analog of the world line, swept by the string in its motion. Spacetime
coordinates appear as {\em fields} on this two dimensional surface, which
is usually assumed to be compact. We will consider bosonic strings,
compactified on a $d$ dimensional torus, $\real^d$ quotiented by an abelian
infinite group (a lattice) $\Gamma$ generated by $d$ generators $e_i$. On
the generators of $\Gamma$ we define an inner product which provides a
metric (of Euclidean signature) on the torus ${\cal T}_d$:
\be
\vev{e_i,e_j}\equiv g_{ij} \ \ . \label{metric}
\ee

The dual lattice $\tilde\Gamma$ is spanned by the basis $e^i$
with (we implicitly complexify $\Gamma$ and extend the product):
\be
\vev{e^i,e_j}=\delta^i_j \ \ .
\ee
The inner products of the $e^i$'s define a metric which is the inverse of
$g_{ij}$, that is:
\be
\vev{e^i,e^j}\equiv g^{ij} \ \ .
\ee
Notice that, if all of the $e_i$ are quantities of order $R$ (we take
Planck's length to be unit unless otherwise stated), with $\det g$ of order
$R^d$, then the `size' of the dual lattice is a quantity of order $1/R$. In
this sense, if to a given lattice corresponds a large universe, to its dual
it will correspond a small one, the dual torus $\tilde{\cal
T}_d$\footnote{Strictly speaking however this conclusion is only valid only
in the absence of torsion (introduced below) in the action
\cite{DouglasHull}.}.

Classically the  string is described by a two dimensional nonlinear $\sigma$
model, whose fundamental objects are the Fubini--Veneziano fields, which,
for the case of a closed string are:
\be
X^i(\tau,\sigma)=x^i+g^{ij}p_i\tau+g^{ij}w_i\sigma
+\sum_{k\neq0}\frac1{ik}~\alpha^{(\pm)i}_k~e^{ik(\tau\pm\sigma)} \ \ ,
\label{FubVen}
\ee
where $x$ represents the centre of mass of the string, $p$ its momentum and
$w$ is the winding number, the number of times the string wraps around the
direction defined by the $e_i$. Notice that, since the space is compact,
the momentum is quantized, and in fact it must be $p\in\tilde\Gamma$, while
the winding number must belong to the dual lattice $w\in\Gamma$. If the
size of the target space is extremely large, then the momentum will have a
spectrum with very close eigenvalues, a nearly continuous spectrum, while
the windings will have values far apart. But apart from these scale
considerations, the role of $p$ and $w$ in \eqn{FubVen} is symmetric. In
the following we will concentrate on the zero modes of the string, mostly
ignoring the oscillator modes. These are internal excitations of the other
string, and are not sensible to the target space in which the strings live,
and will therefore play an indirect role for the structure of spacetime.
Moreover, the oscillators describe excitations which are starting at the
Planck mass, while most of our considerations relate to the low energy
sector of the theory. however thay cannot be ignore altogether, as they
reveal the ``string'' character of the theory.

We have therefore a nonlinear $\sigma$
model described by the action:
\be
S=\frac 1{4\pi} \int d\sigma d\tau \sqrt{\eta}
\eta^{\alpha\beta}\del_\alpha X^i g_{ij} \del_\beta x^j +
\varepsilon^{\alpha\beta} b_{ij}\del_\alpha X^i\del_\beta X^j \ \ ,
\ee
where $\eta$ is the world sheet two dimensional metric, $\varepsilon$ is
the antisymmetric tensor with $\varepsilon_{12}=1$, $G$ is the metric
defined in \eqn{metric}, and $b$ is an antisymmetric tensor which represent
the `torsion' of the string.

We can perform a chiral decompositions of the $X$'s defining:
\be
X^i_\pm(\tau\pm\sigma)=x^i_\pm+g^{ij}p_j^\pm(\tau\pm\sigma)
+\sum_{k\neq0}\frac1{ik}~\alpha^{(\pm)i}_k~e^{ik(\tau\pm\sigma)} \ \ .
\label{chiralmultfields} \label{FubVenpm}
\ee
The zero modes $x^i_\pm$ (the centre of mass coordinates of the string) and
the (centre of mass) momenta $p_i^\pm=2\pi p_i\pm(g-\mp b)_{ij} w^j$ are
canonically conjugate variables,
\be
[x^i_\pm,p_i^\pm]=-i\delta_i^j
\label{cancommmomx}
\ee
with all other commutators vanishing. The left-right momenta are
\be
p^\pm_i=\mbox{$\frac1{\sqrt2}$}\left(p_i\pm\langle e_i,w\rangle
\right)
\label{momlattice}
\ee
The $p^\pm$'s belong to the lattice:
\be
\Lambda=\tilde\Gamma\oplus\Gamma
\ee
We can therefore define the fields $X=X_++X_-$, and we may equally well
define $\tilde X\equiv X_+-X_-$, whose zero mode we will indicate as
$\tilde x$.

Exchange of a lattice with its dual is a symmetry called
T-duality \cite{tduality,GPR}. It corresponds to an exchange of the momentum
quantum number with the winding, and of the zero mode corresponding to $x$,
the position of the centre of mass
of the string with its dual $\tilde x$. This is a symmetry of the
Hamiltonian:
\bea
H&=&\frac 1{2} \left( (2\pi)^2 p_ig^{ij}p_j + w^i(g-bg^{-1}b)_{ij}w^j +4\pi
w^ib_{ik}g^{kj}p_j\right)\nn\\&& +\sum_{k>0}g_{ij}~\alpha_{-k}^{(+)i}
\alpha_k^{(+)j}+\sum_{k>0}g_{ij}~
\alpha_{-k}^{(-)i}\alpha_k^{(-)j}-\frac d{12}
\\ &=&\frac 12 (p_+^2+p_-^2) +\mbox{\rm Oscillators}
-\frac d{12}\label{spectrum}
\eea
with the term $-\frac d{12}$ due to normal ordering pon quantization.
Since momenta and windings belong to a lattice the spectrum is
discrete.

Two target spaces related by a T-dual transformation are indistinguishable
at low energies. This can be seen heuristically as follows
\cite{BrandenbergerVafa}: in ordinary quantum mechanics position is just a
derived concept, as the Fourier transform of momentum spaces. In a string
theory is possible however to consider winding (and its eigenstates) rather
than momentum. If the compactification radius is of the order one, the two
choices are equivalent, but for a very large radius the eigenvalues of
momentum are nearly continuous, while the ones of winding are far apart,
the first one above zero being at a very large energy. It is therefore
difficult to make ``localized wave packets'' with the Fourier transform of
winding. Conversely, with a small radius of compactification, it is the
winding which gives the possibility to create localized wave packets.

In the torsionless
case, $b=0$, this corresponds to an exchange of $g$ with its inverse
$g^{-1}$, and the change of size of the target space in which the radius $R\to
1/R$. In the presence of torsion the exchange is $g^{-1}\leftrightarrow
g-bg^{-1}b$ and $bg^-1\leftrightarrow -g^{-1}b$, and it depends crucially
on the values of the $b_{ij}$. In the toroidal case it is possible to
exchange only some of the generators of the lattice with their duals,
giving rise to a group of factorized T-dualities.

The full group of symmetry is even larger: it is in fact $O(d,d,\inte)$
\cite{Narain,GPR}, generated from three kinds of transformations:
\begin{itemize}

\item[-] The factorized dualities we have already discussed.

\item[-] The changes of base of the lattices, made via a matrix which
belongs to $G(d,\inte)$, the group of integer valued matrices of unit
determinant.

\item[-] The transformation $b_{ij}\to b_{ij}+c_{ij}$ with $c$ an
antisymmetric tensor with integer entries.

\end{itemize}
There is a further $\inte_2$ symmetry obtained exchanging $\sigma$ and
$\tau$ on the world sheet, this last symmetry does not affect the
target space.

The theory has also two continuous symmetries:
\begin{itemize}
\item Target Space reparametrization:
\be
X_\pm(z_\pm)\to X_\pm(z_\pm)+\delta X_\pm(z_\pm)
\ee
\item World sheet conformal invariance, represented by two Virasoro algebras:
\bea
\left[L_k^\pm,L_m^\pm\right]&=&(k-m)L_{k+m}^\pm+\mbox{$\frac
c{12}$}\left(k^3-k\right)\delta_{k+m,0} \nn\\
\left[L_k^-,L_m^+\right]&=& 0
\ ,
\label{vircomm}\eea
\end{itemize}
both these symmetries play a crucial role in the theory.

Given this scenario we want to construct the {\em Noncommutative Geometry}
of interacting strings. We will therefore construct, in the spirit of
Connes, a spectral triple, the Fr\"ohlich-Gaw\c edzki Spectral Triple
\cite{FG}

Let us first construct the Hilbert Space of String states.
Upon first quantization the oscillator modes become creation and
annihilation operators:
\be
\left[\alpha^{(\pm)i}_k,\alpha^{(\pm)j}_m\right]=kg^{ij}
\delta_{k+m,0} \label{creannalg}
\ee
while the zero modes have the usual commutation relations \eqn{cancommmomx}.
The Hilbert Space of (excited) string states therefore is:
\be
{\cal H}=L_2(sp(T^d))^\Gamma\otimes{\cal F}^+\otimes{\cal F}^-
\ee
where $L_2(sp(T^d))^\Gamma$ (spinors on $T^d$) is a set of spinors for each
winding sector, labelled by the lattice $\Gamma$. These are the so called
`tachyon states', although, depending on the actual string theory at hand,
they may not be tachyons (and hopefully they are not, as in superstring
theory). The spaces ${\cal F}^\pm$ are the Fock spaces of higher
excitations (graviton, dilaton etc.) acted upon by the oscillator creation
and annihilation modes. They represent the internal excitations of the
strings and have an indirect effect on spacetime, which is described by the
zero modes.

The description of interacting strings is done via the insertion on the
world--sheet of {\em Vertex Operators}. The fundamental operator is
the``tachyon vertex operator''
\be
V_{q^\pm}(z_\pm)=~:e^{-iq_i^\pm X_\pm^i(\tau\pm\sigma)}:
\ee
where $:\cdot :$ represents normal ordering obtained putting creators to
the left of annihilators:
\bea
:\alpha_k^{(\pm)i}\alpha_m^{(\pm)j}:&=&\alpha_k^{(\pm)i}
\alpha_m^{(\pm)j}~~\mbox{for}~~k<m\\
&=&\alpha_m^{(\pm)j}\alpha_k^{(\pm)i}~~\mbox{for}~~k>m
\label{wickorder}\eea
and $x_\pm^i$ to the left of $p_i^\pm$. The tachyon vertex operator
represents the insertion on the world sheet of a ground state (tachyon) of
a given momentum. higher states (the dilaton, graviton etc.) are obtained
acting with the appropriate combination of creation operators.

Vertex Operator Algebras have a distinguished place in mathematics, they
have connections with modular functions, Monster Group, Lie Algebras and
they are well reviewed in several publications, among wich \cite{VOA}. I
have no room to describe the beautiful mathematical intricacies of the
theory, for most of our purposes in fact vertex operators will just be
operators on ${\cal H}$.

One of the aspects of Vertex Operator Algebras which is important in this
context is the {\em Operator--State correspondence.}  We can put the
generic vertex operator:
\be
V(z_+,z_-)_\psi=:i\,V_{q^+q^-}(z_+,z_-)\mbox{$\prod_j\frac{r_i^{(j)+}}
{(n_j-1)!}\,\partial_{z_+}^{n_j}X_+^i\prod_k\frac{r_j^{(k)-}}
{(m_k-1)!}\,\partial_{z_-}^{m_k}X_-^j$}: \label{highspin}
\ee
in correspondence with the state:
\be
|\psi\rangle=|q^+;q^-\rangle\otimes\mbox{$\prod_j$}\,r_i^{(j)+}
\alpha_{-n_j}^{(+)i}|0\rangle_+\otimes\mbox{$\prod_k$}\,r_j^{(k)-}
\alpha_{-m_k}^{(-)j}|0\rangle_- \label{homstate}
\ee
of $\cal H$, where $(q^+,q^-),(r^+,r^-)\in\Gamma\oplus\Gamma^*$.

We thus have the second element of the spectral triple, an algebra of
vertex operator. A warning however: a vertex operator algebra  (in the
common use of the term) is {\em not a $C^*$ algebra}. In general vertex
operators are not even bounded operators! The problem stems from the fact
that vertex operators are not defined at coinciding points giving rise to
nontrivial {\sl Operator Product Expansions} \cite{GSW}. One can do two
things to regularize the theory: {\em smear} the vertex operators
\cite{FG,FGR}:
\be
V(\psi,f)=\int dz V_\psi(z)f(z)
\ee
but this not always cures the problem, as discussed for example in
\cite{CS}. An alternative is to consider {\em truncated Vertex operators}:
\be
V_{q^\pm}^{N}(z)={\cal N}_N\prod_{n=0}^N W_n
\ee
where $W_0$ contains the zero modes $x$ and $p$, while the $W_n$'s $(n\neq
0)$ involve only the $n^{th}$ oscillator modes $\alpha_{n}^{(\pm)}$ and
$\alpha_{-n}^{(\pm)}$. This is equivalent to an ultraviolet cutoff on the
world sheet, a standard practice in string theory to avoid the infinities
arising from the product of operators at coincident points. At the end one
considers $N\to\infty$.

It is however fair to say that, at present, the rigourous definition of a
$C^*$-algebra of operators representing interacting strings is (at least to
my mind and my knowledge) still an open problem. We have the tachyon
operators and the higher spin state \eqn{highspin}, and one should
regularize them, and create an algebra with the appropriate completion. It
is in a sense like attempting to construct $C(\real)$ from the knowledge of
plane waves $e^{ipx}$. The general idea is present but many (crucial)
details have to be filled. This is an area in which the collaboration of
mathematicians would be of paramount importance. In the following we will
indicate with the generic term vertex operator algebra a proper completion
of the regularized operators.

We can easily recognize the two fundamental symmetries of the theory in the
vertex operator algebra. As we said the tachyon operators are in a sense a
``Fourier'' or plane waves basis on the space of conformal field
configurations. The tachyon states are highest weight states of the level a
pair of $u(1)_+^d\oplus u(1)_-^d$ current algebra
\eqn{creannalg}, so that the entire Hilbert space can be built up from the
actions of the $\alpha_k$'s for $k<0$ on these states. This current algebra
represents the target space reparametrization symmetry of the string
theory. On the other side, the two Virasoro algebras which represent the
world sheet conformal invariance have irreducible representation whose
highest weights grade the Hilbert space $\cal H$. The Virasoro operators
in the present case are
$L_k^\pm=\frac12\sum_{m\in\inte}g_{ij}:\alpha_m^{(\pm)i}
\alpha_{k-m}^{(\pm)j}:$, with $\alpha_0^{(\pm)i}\equiv
g^{ij}p_j^\pm$. They generate a representation of the Virasoro algebra
\eqn{vircomm} of central charge $d$. The grading is defined on the
subspaces ${\cal H}_{\Delta_q}\subset\cal H$ of states \eqn{homstate} which
are highest weight vectors,
\be
L_0^\pm|\psi\rangle=\Delta_q^\pm|\psi\rangle~~~~~~,~~~~~~
L_k^\pm|\psi\rangle=0~~~\forall k>0 \ ,
\label{highestwt}\ee
where $\Delta^+_q=\frac12g^{ij}q_i^+q_j^++\sum_jn_j$ and
$\Delta_q^-=\frac12g^{ij}q_i^-q_j^-+\sum_km_k$. The corresponding
operator-valued distributions \eqn{highspin} are called primary fields.

The last element to complete the spectral triple is the Dirac operator. We
have not one but {\em two} natural Dirac Operators:
\be
D^\pm=\gamma_i^\pm\alpha^i_\pm \ \alpha^i_\pm=-i\del_\pm X^i_\pm
\ee
These two operators generate target space reparametrization of $X_\pm$.
Moreover it can be seen that they they are square roots of the
Laplace--Beltrami operator. They are also naturally related to the other
symmetry of the string theory, in fact worldsheet conformal symmetry has
the conserved stress energy:
\be
T^\pm(z_\pm)=-\frac 12 :D^\pm(z_\pm)^2:~ =\sum_k L_k^\pm z_\pm^{-k-2}
\ee
In analogy with the $X$ and $\tilde X$ we can define:
\be
D=D^++D^- \ \ ; \ \tilde D=D^+-D^-
\ee
The spectral triple ${\cal T}$ of string geometry therefore is:
\be
{\cal A} \ \ {\cal H} \ \ D
\ee

One can ask now what happened to ordinary spacetime? Spacetime emerges as a
``subtriple'' ${\cal T}_0$, that is, a spectral triple with a subalgebra, a
subspace of Hilbert space, and an operator which is the reduction of the
Dirac operator on the subspace:
\be
{\cal A}_0 \ \ {\cal H}_0 \ \
\slash\partial
\ee
In order to construct the low energy subtriple we  first have to project
out all of the oscillator modes to obtain ${\cal A}_0$ and ${\cal H}_0$.
The rationale behind this is that, since the excited oscillators start at
the Planck mass, and this is much larger than `ordinary' space time
energies, we have to isolate the modes of the string which will be
accessible at low energies. This is still not sufficient however, as, in
the case of large uncompactified directions, the modes associated to the
winding are also highly energetic. We therefore choose:
\be
\begin{array}{cccc}
C(\real^d):& f\in{\cal A}_0:& [\tilde D,f]=0 & \mbox{\tiny commutant of
$\tilde D$}\\
L_2(T^d,sp):& \psi\in{\cal H}_0: & \tilde D\psi=0& \mbox{\tiny kernel of
$\tilde D$}
\end{array}
\ee

It is easy to connect ${\cal A}_0$ with the algebra of complex valued
function on spacetime, it is sufficient to notice that it is constructed
from the (commutative) `vertex operator' $e^{ipx_0}$. Here we encounter
the already mentioned problems of the appropriate completion in order to
obtain a well defined \cstar. The essence of T-duality lies in the
relatively simple observation the instead of $\tilde D$ we could have
chosen $D$ as well. In this case we would have obtained the triple
pertaining to the torus whose coordinated are the $\tilde x$, that is the
T-dual torus $\tilde {\cal T}_0$, with all the radii of compactifications
inverted:
\be
D^\pm\to D^\mp \ \ \
D\leftrightarrow\tilde D
\ee

That the full theory is invariant under his change is ensured by the
observation that this transformation is a {\em gauge} transformation. In
fact there are (many) $u\in{\cal A}$ unitary such that:
\be uDu^{-1}=\tilde D\ee
For example:
\be
u=e^{i{\cal G}_\pm}
\ee
with
\be {\cal
G}_\chi=\int\frac{dz_+~dz_-}{4\pi
z_+z_-}~\left(\chi^a_{+,i}[X]\,J^{+(i)}_a(z_+)
+\chi^a_{-,i}[X]\,J^{-(i)}_a(z_-)\right)f_S(z_+,z_-)
\label{gaugegen}
\ee
where $a=\pm$, $J_a^{\pm(i)}=:e^{aik_j^{(i)}X^j_\pm}:$, $\chi_\pm$ are
sections of the spin bundle and $f_S$ is a smearing function.

This T-duality is a however a gauge transformation only in the full
FG-triple. When this is projected to the subtriple so to give a {\em
commutative} spacetime, in general the process will give rise to very
different spacetimes. We can in fact consider T-duality to be the
commutativity of the following diagram:
\be
{\begin{array}{rrl} {\cal
T}_D&{\buildrel u\over\longrightarrow}&{\cal T}_{\tilde D}\cong{\cal
T}_D\\{\scriptstyle{\cal P}_0}\downarrow&
&\downarrow{\scriptstyle\tilde{\cal P}_0}\\{\cal T}_0&{\buildrel
T_0\over\longrightarrow}&\tilde{\cal T}_0\end{array}}
\ee
The operation $T_0$ is what we call T-Duality, and from the previous
discussion it is clear that it is just the low energy projection of a gauge
transformation. All of the remaining $O(d,d,\inte)$ dualities can be
obtained in the same way, as gauge transformations \cite{lizziszabocmp}.

There are many more inner automorphisms, {\em gauge transformations}, which
project down to non trivial transformations. Defining the
currents:
\be
J_\pm^i(\tau\pm\sigma)=\partial_\pm
X_\pm^i(\tau\pm\sigma)=\sum_{k=-\infty}^\infty
\alpha_k^{(\pm)i}~e^{ik(\tau\pm\sigma)}
\label{currentexp}
\ee
 a general spacetime coordinate transformation $X\to\xi(X)$, with
$\xi(X)$ a local section of ${\rm spin}(T^n)$, is generated by ${\cal
G}_\chi={\cal G}_\xi$ with
\be
{\cal
G}_\xi=\int\frac{dz_+~dz_-}{4\pi
z_+z_-}~\xi_i(X)\left(J_+^i(z_+)+J_-^i(z_-)\right)f_S(z_+,z_-)
\label{gencoordauto}\ee
The means that the also the diffeomorphisms of the (low energy) target
space are gauge transformations of the full spectral triple. The inner
automorphisms project down to outer automorphisms of spacetime. This is, in
my opinion, one of the best justifications of the often heard statement
that ``general relativity is a gauge theory". We can see also a glimpse of
an huge group of symmetries, which when projected down connects different
low energy theories.

If we try then to uncover the structure of spacetime at higher energies we
would have to consider momentum and winding modes on a par. This will be
relevant when the radius of compactification is comparable with Planck's
length, as in this case is not possible to ignore the former over the
latter. We will however limit ourselves for the time being to the tachyonic
case. Nevertheless the oscillators (at least the lower ones) do play an
important role. Consider therefore tachyon vertex operators, for which we
only excite the first $N$ oscillators (for the basis $e^i_\pm$ of
$\gamma\oplus\Gamma^*$). The commutation relation among the elementary
operators are:
\begin{eqnarray}
V^N_{e^i_\pm}({z_\pm}_i)\,V^N_{e^j_\mp}({z_\mp}_j)&=&V^N_{e^j_\mp}({z_\mp}_j)
\,V^N_{e^i_\pm}({z_\pm}_i)\label{VVpm}\nn\\
V^N_{e^i_\pm}({z_\pm}_i)\,{V^N_{e^i_\pm}({z_\pm}_i)}^\dagger&=&
{V^N_{e^i_\pm}({z_\pm}_i)}^\dagger
\,V^N_{e^i_\pm}({z_\pm}_i)~=~\unit\label{VV*}\nn\\
V^N_{e^i_\pm}({z_\pm}_i)\, V^N_{e^j_\pm}({z_\pm}_j)&=&e^{2\pi
i{\omega_N}_\pm^{ij}}~V^N_{e^j_\pm}({z_\pm}_j)\,V^N_{e^i_\pm}({z_\pm}_i)
{}~~~~,~~~~i\neq j \label{nctorop}
\end{eqnarray}
where the ${z_\pm}^i$ are distinct points, and
\be
{\omega_N}_\pm^{ij}=\pm
g^{ij}\left(\log\left(\frac{z^\pm_i}{z^\pm_j}\right)-
\sum_{n=1}^N\frac{1}{n}\left(\left(\frac{z^\pm_i}{z^\pm_j}\right)^n
-\left(\frac{z^\pm_j}{z^\pm_i}\right)^n\right)\right)\ \ .
\ee
One can easily recognize in \eqn{nctorop} a {\em noncommutative torus}
structure \cite{Rieffel}. If we enclose more and more oscillators:
\be
\lim_{N\to\infty}{\omega_N}_\pm^{ij}=\omega_\pm^{ij}=\pm\,g^{ij}~{\rm
sgn}({\rm arg}\,z_i^\pm-{\rm arg}\,z_j^\pm)\  i\neq j
\ee

The symmetries of the theory are still present, even in this truncated
version, in fact theories related by $O(d,d,\inte)$ transformations give
rise to Morita equivalent tori \cite{LLS,Rimor}. The commutative case is
recovered in the uncompactified/large compactification radius because when
$R\to\infty$, the off diagonal elements of $g^{ij}\to 0$ and we recover the
commutative torus.

``Turning the NCG crank'' it is also possible to write a {\em Low Energy
Dual Symmetric Action} \cite{LLS}:
\bea
{\cal L}&=&\left(F+{ }^\star F\right)_{ij}\left(F+{ }^\star F\right)^{ij}
\nn\\ &&-i\,\overline{\tilde
\psi}\,\gamma^i\left(\partial_i+ i{\buildrel\leftrightarrow\over
{A_i}}\right)\,\psi-i\,\overline{\psi}\,\tilde \gamma_i\left(\tilde
\partial^i+i
{\buildrel\leftrightarrow\over{\tilde A^i}}\right)\,\tilde \psi
\eea
where the dual field strength is defined:
\bea
F_{ij}&=&\partial_iA_j-\partial_jA_i+i\Bigl[A_i,A_j\Bigr]\nn\\ &&
-g_{ik}\,g_{jl}
\left(\tilde \partial^k\tilde A^l-
\tilde \partial^l\tilde A^k+i\left[\tilde A^k,\tilde A^l\right]\right)
\eea

All of the $O(d,d,\inte)$ transformations, being unitary transformations,
do not change the {\sl spectrum} of $D$. Let us analyse in some details the
transformation which changes the components of the antisymmetric second
rank tensor $b_{ij}$ by the addition of an arbitrary, integer valued,
constant matrix. Altough this transformation does not change the lattice
$\Gamma$, it will change the momenta conjugated to the zero modes of $X$
and $\tilde X$. In particular, in the spectrum \eqn{spectrum}, the relative
contribution of the momenta (represented by the first term,) with respect
to the windings, and the mixed term will change.  Choosing the components
of the antisymmetric tensor $b$ arbitrarily large, we can make the
contribution of the second and third term arbitrarily large. We have
therefore concentrated the lowest eigenvalues of the Hamiltonian in the
momentum part. The low energy spectrum is made only of the momentum
eigenvalues. The lattice is still the same, but the strings are extremely
twisted, and we have transferred the lowest eigenvalues of the energy from
winding to momentum. Roughly speaking, a low energy strings, which in the
original (small radius) lattice had a combination of momentum and winding,
will now be twisted in such a way that it will appear to have just
momentum, it is like the lattice ``repeats itself over and over''.

Again, as in the case of the of the $R\leftrightarrow 1/R$ symmetry, we
have to ask ourselves `what is position'? `How is it measured'? And using
the same heuristic arguments of \cite{BrandenbergerVafa}, we can think of
making wave packets using superpositions of the eigenvalues of the momentum
In the case of large torsion the eigenvalues of momentum are continuous for
all practical purposes, therefore the superposition will have the character
of a uncompactified space, rather than a string moving on a lattice. And
this will be the situation until energies in which the new eigenvalues
(coming from windings or the oscillatory modes) start to play a role.

Let us briefly discuss the role of the classical configuration space in
ordinary quantum mechanics in the language and formalism of noncommutative
geometry. We will be very brief and refer to \cite{size} for further
details and references. Consider a purely quantum observer, that is a set
of operators which form an algebra. For example bounded operators
constructed from $p$ and $x$. The information on the topology of $M$, the
manifold on which the motion is happening, can be recovered in the
programme of noncommutative geometry by considering the algebra of {\em
position operator}, that is, the algebra of continuous\footnote{In the
following we will consider $M$ compact, therefore continuous functions are
bounded as well.}, complex valued, functions on $M$, seen as operators on
$L^2(M)$, with a norm given by the maximum of the modulus of the function.
This is a simple application of the Gel'fand--Naimark theorem.

We will consider the configuration space of a quantum mechanical space
therefore not as a set of points, but rather as an abelian $C^*$-algebra.
The Hilbert space could also be easily constructed a posteriori by giving a
sesquilinear form (a scalar product) on the algebra, and completing it
under the norm given by this product. Other choice for the Hilbert space
are possible, a relevant one for instance is the space of spinors. A
quantum observer will have at his disposal, among the bounded operators on
the Hilbert space, an abelian subalgebra ${\cal A}_0$ which he will
identify with the continuous function on his space.

The ``size'' of this configuration space is given by the Dirac operator via
Connes' distance formula \cite{book}:
\be
d(x,y)=\sup_{||[D,a]||\leq 1}\left|a(x)-a(y)\right| \ \ a\in{\cal A}_0\ \ .
\label{distances}
\ee

Noncommutative geometry equips our quantum observer with a series of tools
suited to him: algebras of operators, traces etc. In the commutative case
these tools reconstruct the usual differential geometry,  but they can be
used in the noncommutative case as well. If we are in a commutative case,
the quantum observer has therefore at his disposal an algebra of
observables, in this algebra he recognizes an abelian subalgebra, that he
calls the space on which he lives, and with formula
\eqn{distances} he calculates distances, metric etc.

In string theory, spacetime, as described for example by \eqn{nctorop},the
quantum observer finds himself on a {\em noncommutative space}. That is,
among his set of quantum observable he does not identify an abelian algebra
giving him the configuration space, he can however define some sort of
``noncommutative'' space, to which it corresponds a noncommutative algebra.

%%%%%%%%%%%%%%%%%%%%%%%%%%%%%%%%%%%%%%%%%%%%%%%%%%%%%%%%%%%%%%%%%%%%%%%%%%%%%
To specify the meaning  of ``low energy'' we will resort to the spectral
action principle \cite{ChamsConnes}, and will argue that the meaning of
low energy means
a theory in which only the low part of the spectrum of $D$ plays a role.
This is possible because in the framework of Noncommutative
geometry one constructs a {\em spectral geometry}, in which the information
is stored in the spectrum of $D$. And low energy refers to an action in
which only the lower part of the spectrum is excited.

The spectral action principle is based on the covariant Dirac operator, and
on the variation of its eigenvalues. The action must be read in a Wilson
renormalization scheme sense, and it depends on an ultraviolet cutoff
$m_0$:
\be
S_{m_0}=\Tr\chi\left({D_A^2\over m_0^2}\right) \label{specac}
\ee
where $D_A$ is the covariant Dirac operator and $\chi(x)$ is a function
which is 1 for $x\leq 1$ and then goes rapidly to zero (some smoothened
characteristic function). The action \eqn{specac} effectively counts the
eigenvalues of the covariant Dirac operator up to the cutoff. Considering,
in fact, the eigenvalues of $D_A$ as sequences of numbers, and these
sequences as dynamical variables of euclidean gravity, the spectral action
is then the action of ``general relativity'' in this space
\cite{LandiRovelli}. The trace in the action can be calculated using known
heath kernel techniques \cite{ChamsConnes}, and the resulting theory
contains a cosmological constant, the Einstein--Hilbert and Yang--Mills
actions, plus some terms quadratic in the Riemann tensor.
%Chamseddine
%\cite{Chamseddine} has used the Dirac operator in the spectral action
%principle and shown that they lead to the low energy effective string
%action.

 What is important in the present context is the {\em spectral
principle}, that is, the starting point is the spectrum of an operator, and
its variations as the backgrounds fields (the one--form $A$ in this case)
change. One can ask, in fact, what is the role of the algebra in the
spectral action, as the latter depends just on the trace of the Dirac
operator. Of course the role of the algebra is in the fact that in
\eqn{specac} appears the {\em covariant} Dirac operator. And the form
$A=\sum a_i[D,b_i]$ depends on the algebra chosen. Let us now apply these
considerations to the Fr\"ohlich-Gaw\c edzki spectral triple.

The spectrum of $D$ and $\tilde D$, or of any operator obtained from them
with an $O(d,d,\inte)$ unitary transformation, are the same. Let us call
$D$ for convenience the one for which the lowest eigenvalues are the one
relative to momentum. Here by lowest we mean the ones which are lower than
the energy of the oscillatory modes (of the order of the Planck mass
$m_p$). If the cutoff $m_0$ is lower than $m_p$, the cutoff function $\chi$
causes the projection of the operator on the Hilbert space ${\cal H}_0$.
Elements of the algebra which commutes with $D$ (such as the elements of
$\tilde{\cal A}$) will not contribute to the variations of the action, and
will therefore be unobservable. This algebra can be constructed as the
commutant of the T-dual operator $\tilde D$. This means that the winding
modes degrees of freedom are unobservable. Since the Dirac operator has a
near continuous spectrum, the tachyonic, low energy, algebra is spanned by
operators of the kind
\be
V_p=e^{ipx} \ \ ,
\ee
can be considered the Fourier modes describing
an uncompactified space.

In fact, a quantum observer with a spectral action, will be able to measure
(in the form of fields, potentials etc.) only the elements of the algebra
which give low energy perturbations of the lowest eigenvalues of $D$,
always with the assumption of the cutoff $m_0<m_p$ so that oscillatory
modes do not play a role. This is the abelian algebra of functions on some
space time. If, as we have seen, there are many low eigenvalues, the
observer will experience an effectively decompactified space time. The
algebra which he will measure will be composed of the operators which will
create low energy perturbation to $D$. At this point we have to make the
sole assumption that $D$ has a spectrum with several small eigenvalues. In
this way the quantum observer will experience a (nearly) continuous
spectrum of the momentum, the sign of an uncompactified space.

The strings could still be seen as compactified on a ``small'' lattice, but
the presence of a very large torsion term $b$ has drastically changed the
operator content of the theory, and this has rendered space effectively
uncompactified.

%%%%%%%%%%%%%%%%%%%%%%%%%%%%%%%%%%%%%%%%%%%%%%%%%%%%%%%%%%%%%%%%%%%%%%%%%%%%%%%%5

\subsubsection*{ Conclusions}

In both String Theory and Noncommutative Geometry, the interaction between
physics and mathematics has been very fruitful, but he mathematics used in
string theory has been essentially ``classical" differential geometry. In
this talk I tried to give an impressionistic way on how the mathematics
well suited to describe strings in the high energy regime (which is proper
to them) should be some sort of noncommutative geometry. While from the
physical point of view some (initial) result are already to be seen: {\em
duality, gauge transformations \ldots}, from the mathematical point of view
the structures to use are still in need of proper definitions.

A proper mathematical sharpening of the tools is necessary not so much for
abstract mathematical rigour, but to help uncover the beauty which lies
behind such a rich structure.

\bigskip

\noindent
{\bf Acknowledgements.} \\ I would like G.~Landi and R.J.~Szabo for a very
fruitful collaboration, F.~Scheck, H.~Upmeier and W.~Werner for organizing
the Arbeitstagung and inviting me to participate in it, and principally
{\em all} of the participants for making the workshop a most interesting
and enjoyable one!

\bibliographystyle{unsrt}

% ----------------------------------------------------------------
\end{document}